\documentclass[twocolumn,preprintnumbers,amsmath,amssymb]{revtex4}

\usepackage{dcolumn}
\usepackage{amstext}
\usepackage{amsmath}
\usepackage{amssymb}
\usepackage[pdftex]{graphicx}
\usepackage{subfigure}
\usepackage[T1,OT1]{fontenc} 

\begin{document}
\title{Effective elasticity of a flexible filament bound to a deformable cylindrical surface}
\author{An{\fontencoding{T1}\selectfont\dj}ela \v{S}ari\'c, Josep C. P\`amies and Angelo Cacciuto}
\affiliation{Department of Chemistry, Columbia University\\ 3000 Broadway, MC 3123\\New York, NY 10027 }
\renewcommand{\today}{November 2, 1994}

\begin{abstract}
We use numerical simulations to show how a fully flexible filament binding to a deformable
cylindrical surface may acquire a macroscopic persistence length and a helical conformation. This is a result of the
nontrivial elastic response to deformations of elastic sheets.
We find that the filament's helical pitch is completely determined by the mechanical
properties of the surface, and can be tuned by varying the filament binding energy. We
propose simple scaling arguments to understand the physical mechanism  behind this phenomenon and
present a phase diagram indicating under what conditions one should expect a fully flexible chain to behave as a 
helical semi-flexible filament. Finally, we discuss the implications of our results.
\end{abstract}
\maketitle

A ubiquitous geometrical state filaments arrange into is the helix. Apart from some synthetic polymers~\cite{nakano} 
and biological filaments such as ds-DNA and actin filaments
which spontaneously develop a helical conformation due to their inherent chemical structure~\cite{alberts}, 
helicity can also appear when a filament is bound to a cylindrical surface. This phenomenon  can be observed across all length scales:
from vine wrapped around trees, to DNA on carbon nanotubes~\cite{Zheng}.

Although in several instances it is believed that what leads to the helicity of a filament is either a specific property
of the filament or the specific interactions between the filament and the underlying
 surface~\cite{shaw,milligan,Balavoine}, there is evidence that semiflexible polymers binding non-specifically to 
 cylindrical surfaces can spontaneously 
develop helical conformations.   The arrangement of cellulose microfibrils in the plant cell wall~\cite{neville} 
is a nice example of it. Recently it has been suggested that the helix is the preferred conformation of semiflexible polymers when 
generically bound to the surface of an infinitely long cylinder, provided the cylinder's radius is sufficiently large~\cite{srebnik1}. 

One particular aspect of the  problem that has not been studied and could be of great relevance, concerns the role of the deformability of the underlying surface.
This property is inherent to biological materials, and the dynamical interplay between protein filaments and the soft cell membrane has been shown to be crucial 
in several biological processes~\cite{alberts}. In fact,  semiflexible biopolymers such as microtubules and actin filaments 
not only provide the cell with a highly dynamical scaffolding  that regulates its shape,
 but they also mediate important extracellular interactions. 
Cell division~\cite{adams} and cell crawling~\cite{bray} are two dramatic examples of it.

Here, we explicitly consider the role of the surface deformability and predict that new phenomenological behavior arises when a filament is bound to it. 
We show how even a fully flexible filament, weakly bound to a soft tubular sheet, can acquire an effective, tunable helicity 
and a large bending rigidity. The physical reasons behind it are quite general, are applicable to arbitrary geometries,
 and can be understood by analyzing
 the nontrivial mechanical response of  elastic sheets to local deformations.

Unlike fluid interfaces, the deformation of an elastic sheet away from its equilibrium shape comes with a bending and stretching energy cost.
It is easy to show~\cite{landau,witten} that the ratio between stretching and bending for an arbitrary deformation of amplitude $h$ on a surface of thickness $t$
scales as $E_s/E_b \sim (h/t)^2$. Therefore, for sufficiently thin sheets, bending is the preferred mode of
deformation. This has a profound effect on the way elastic surfaces respond to deformations as the only stretch-free deformation involves
uniaxial bending. Skin wrinkling under applied stress~\cite{maha1,skin} and stress focusing via d-cone formation of crumpled paper~\cite{witten}
are two beautiful examples of this phenomenon.
 
Using simple scaling arguments it is possible to estimate the extent of the deformation, $l_p$, that arises
when an indentation of amplitude $h$ is imposed on a narrow elastic sheet of width $D$. Following reference~\cite{witten}, the bending and stretching energies associated with this deformation
scale as $E_b\sim K_b\left(h/D^{2}\right)^{2}Dl_p$ and $E_s\sim K_s\left(h^{2}/l_p^{2}\right)^{2}Dl_p$, where $K_b$ and $K_s$ are the bending and stretching
constants respectively. The balance between the two terms gives
$l_p\sim D h^{\frac{1}{2}}\left(K_s/K_b\right)^{\frac{1}{4}}$. A more familiar form of this expression is obtained by plugging
$K_s\sim {\rm Y} t$ and $K_b\sim {\rm Y} t^3$ \cite{landau} (Y is the Young modulus of the surface) to give $l_p\sim D (h/t)^{\frac{1}{2}}$.

This result
can be generalized to a cylindrical surface of radius $R$ (with $D\sim R$) as long as $h \ll R$, and saturates to $l_p\sim R (R/t)^{\frac{1}{2}}$
for thin cylinders and/or large deformations~\cite{pablo,maha2}. Either way, for a fixed cylindrical radius $R$ and indentation $h$,
the extent of the deformation along the axis of the cylinder is set by the ratio between  bending and stretching constants.

In this paper we show how a fully flexible filament that generically binds to a deformable cylindrical surface can acquire a macroscopic bending rigidity
and a specific intrinsic curvature set by the mechanical properties of the surface and the extent of the deformation. The net result is an effective 
semi-flexible chain that wraps around the cylinder with a tunable pitch.
Using a combination of scaling arguments and numerical simulations we show how the characteristic length scale $l_p$
is directly related to the pitch of the helix, and we present a phase diagram showing the transition from a disordered (random walk) to
the helical conformation of the filament as a function of its binding affinity to the surface.

We model the elastic surface via a standard triangulated mesh~\cite{mesh}. The mesh is composed of $N=14960$ nodes arranged
to produce an initial configuration with
perfect hexagonal tessellation. To impose surface self-avoidance we place hard beads in each node of the mesh.
Any two surface beads  interact via a purely repulsive truncated and shifted  Lennard-Jones potential
\begin{equation} \label{LJ}
U_{LJ}=
\begin{cases}
4\epsilon\left[ \left( \dfrac{\sigma}{r}\right)^{12}-\left(\dfrac{\sigma}{r}\right)^{6} + \frac{1}{4}\right] & \text{, $r\leq 2^{1/6} \sigma$}\cr
0 &\text{, $r>2^{1/6} \sigma$}\cr
\end{cases}
\end{equation}
where $r$ is the distance between the centers of two beads, $\sigma$ is their diameter, and
$\epsilon=100 k_{\rm{B}}T$.

We enforce the surface fixed connectivity by linking every bead on the surface to its first neighbors via a harmonic spring potential
\begin{equation} \label{spring}
U_{stretching}=K_s(r-r_B)^2
\end{equation}
Here $K_s$ is the spring constant and  $r$ is the distance between two neighboring beads. $r_B=1.23\sigma$ is the equilibrium bond length, and it is sufficiently short to prevent overlap between any two triangles on the surface even for moderate values of $K_s$.

The bending rigidity of the elastic surface is modeled by a dihedral potential between  adjacent triangles on the mesh:
\begin{equation} \label{dihedral}
U_{bending}=K_b(1+\cos\phi)
\end{equation}
where $\phi$ is the  dihedral angle between opposite vertices of any two triangles sharing an edge and  $K_b$ is the bending constant.

The polymer is constructed as a ``pearl necklace'' with $N_m=20$ monomers of diameter of $\sigma_m=10\sigma$.
Neighboring monomers are connected by harmonic springs as in Eq.~\ref{spring} with the equilibrium bond length $r_M=1.18 \sigma_m$
and spring constant of 120$k_{\rm B}T/\sigma^2$. Polymer self-avoidance is again enforced via the repulsive truncated-shifted Lennard-jones potential
introduced in Eq.~\ref{LJ} with $\sigma\rightarrow\sigma_m$. Note that we do not associate an explicit bending rigidity to the polymer which behaves as a simple self-avoiding random walk
when bound to an infinitely rigid cylinder.

The generic binding between polymer and surface is described by a Morse potential:
\begin{equation} \label{morse}
U_{Morse}=
\begin{cases}
D_0\left( e^{-2\gamma(r-r_{MB})}-2e^{-\gamma(r-r_{MB})}\right) & \text{, $r\leq 10 \sigma$}\cr
0 &\text{, $r>10 \sigma$}\cr
\end{cases}
\end{equation}
where $r$ is the center-to-center distance between a monomer and a surface-bead,  $r_{MB}$ is bead-monomer contact distance $r_{MB}=5.5\sigma$ and
$D_0$ is the binding energy. The interaction cutoff is set to 10$\sigma$ and $\gamma=1.25/\sigma$.

We used the {\sc LAMMPS} molecular dynamics package~\cite{lammps}
with a Nos\'{e}/Hoover thermostat in the $NVT$ ensemble to study the statistical behavior of the system.
Periodic boundary conditions are imposed to make the cylinder effectively 
infinite. No difference was found when using the $NP_zT$ ensemble, with $P_z=0$ ($z$ is aligned along the cylinder's axis).
The timestep size was set to $dt=0.002\tau_0$ ($\tau_0$ is the dimensionless time)
and each simulation was run for a minimum of $5 \cdot 10^{6}$ steps.
The radius of the undeformed cylinder was set to $R=14\sigma$ in all our simulations.

The overall strategy of our numerical work is to perform a statistical analysis of the system
for different values of $K_s$ and $D_0$, and to understand how the configurational properties of the binding polymer
are related to the elastic properties of the templating surface.

Figure 1 shows for a particular value of the membrane bending rigidity the different phases of the polymer  
in terms of the binding constant $D_0$, which regulates the extent of the surface indentation $h$, and the stretching constant $K_s$.
We find a gas phase, an arrested phase and a helical phase.
\begin{figure}[h!]
\center
\includegraphics[width=60mm]{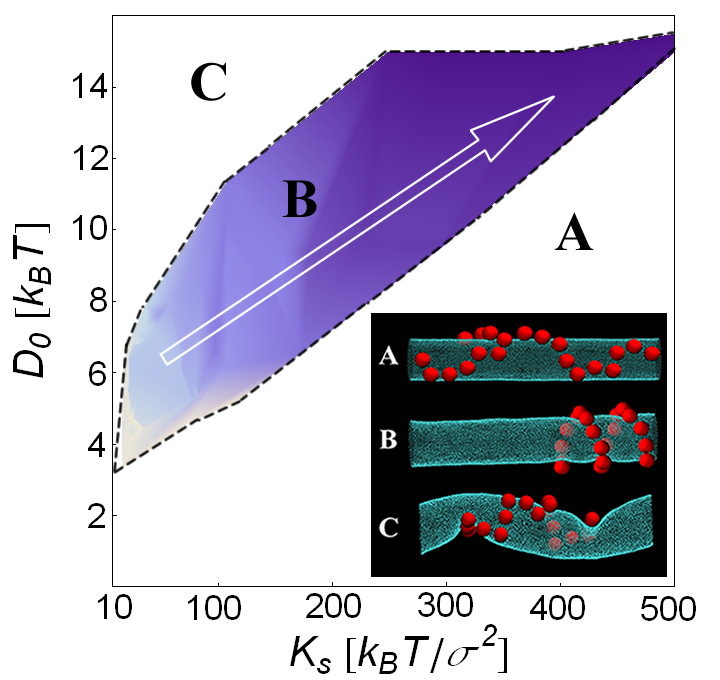}\caption{ Phase diagram of a fully flexible polymer binding to an elastic tubular surface 
for fixed $K_b=150 k_{\rm B}T$. Three phases are shown as a function of $D_0$ and $K_s$ - A: gas, B: helix, C: arrested phase. The direction of the white arrow and the shading in the B phase 
show the helical pitch increase with $K_s$. Dark regions indicate large pitch, whereas light regions represent low pitch. The inset shows three snapshots  
of the chain configurations in the three phases.}
\end{figure}
The behavior of the system in the limit of very large and very small indentations is clear.
In the first case, the cylinder is effectively rigid and does not alter the behavior of the polymer which performs a self-avoiding random walk
over its surface. We indicate this phases as the gas phase.
In the second case, the polymer {acquires non-helical conformations that differ from each other once simulations are repeated 
(under the same conditions) using a different initial configuration. This is indicative that the polymer becomes kinetically trapped 
and we take this as a signature that the system dynamics is becoming glassy.  We call this phase the arrested phase.
The most interesting behavior arises for moderate indentations, where the interplay between
bending and stretching energies of the surface strongly affects the configurations of the polymer, and results in
an interesting helical phase with pitch increasing monotonically with the membrane stretching cost. 

Local deformations caused by each monomer in the helical phase pair-up coherently to generate a smooth surface channel following 
the chain profile. Scaling arguments can be used to estimate the energy cost required to form a channel along  the cylinder axis and one around it.
The total bending energy associate with the axial configuration scales as $E_b^{\parallel}\sim K_b\left({h}/{R^{2}}\right)^{2} l R$,
while that for the transversal configuration has a bending cost $E_b^{\perp}\sim K_b\left({h}/{l_{p}^{2}}\right)^{2} l l_p$,
where $l\sim \sigma_m N_m$ is the contour length of the polymer.

As $l_p$ is typically larger than $R$, $l_p\sim R\sqrt{h/t}$, the bending energy balance favors configurations in which the polymer
wraps around the cylinder to produce ring-like
configurations. However, the  stretching energy 
becomes negligible when the polymer is placed along the cylinder's axis, and grows as
$E_s^{\perp}\sim K_s\left({h}/{l_p}\right)^{4}l l_p$ when it is placed across the axis.
The net result is that when the polymer is bound
to a surface that is easily stretchable, i.e. sufficiently thick, it will spontaneously wrap around its axis. In the limit of an unstretchable, i.e. very thin surface,
the polymer will align with the cylinder axis. The intermediate regime is dominated by helical configurations which represent a balance between the
two tendencies.
By holding $h$ constant and altering the relative weight of bending and stretching energies we can modulate the
pitch of the helix and establish its dependence on the mechanical properties of the membrane.

The angle $\theta$ formed between the axis of the cylinder and the direction of the polymer can be dimensionally related to the two natural length scales of the problem: the axial, $l_p$, and the transversal, $R$
\begin{equation}
\tan(\theta)\sim \left (\frac{R}{l_p}\right )\sim \left(\frac{1}{h^{\frac{1}{2}}\left(K_s/K_b\right)^{\frac{1}{4}}}\right).
\label{thetastretch}
\end{equation}

This functional form has the correct limiting behavior. In the stretching dominated regime $\theta\rightarrow 0$, and in the bending dominated regime
$\theta\rightarrow \pi/2$. It is important to notice that one should be able to modulate the helicity of the polymer by increasing  its binding energy to the surface (i.e. $h$).
However, for sufficiently large values of $h$ the system can become kinetically trapped, or crosses over to the scaling behavior $l_p\sim R^{3/2}/t^{1/2}$~\cite{pablo}, which is independent of $h$. It is therefore clear how variations of $h$ have a weak effect on the pitch of the polymer.

To test our theoretical predictions, we performed a series of numerical simulations  in which we carefully investigated
the dependence of $\theta$ on the membrane stretching rigidity, and on the indentation $h$.
The amplitude of the indentation, $h$,
is tuned by changing the strength of the monomer-bead attraction (binding energy) $D_0$,
and can be estimated by computing the  largest vertical distance among
the surface beads underneath a given monomer.
Fig. 2a shows how $\theta$ depends on $K_s$ for fixed bending rigidity, $K_b$, and indentation, $h$.
The line is a fit to the data obtained  by using the inverse of the functional form in Eq.~\ref{thetastretch}.

Fig. 2b shows how $\theta$ depends on the binding energy $D_0$ which, for fixed $K_s$ and $K_b$, and within the narrow range of values of $D_0$ we explored,
grows linearly with $h$. We repeated the calculation for two different values of $K_s$ and fit the data  with the inverse of Eq.~\ref{thetastretch}.
In both cases Eq.~\ref{thetastretch} appropriately describes the helicity of the polymer in terms of the elastic properties of the membrane.
The inset of Fig. 2a shows the representative snapshots of the polymer conformations for  different values of $K_s$ at constant $h$ and $K_b$.
\begin{figure}[h!]
\center
\subfigure[]{\label{fig:edge-a}\includegraphics[width=60mm]{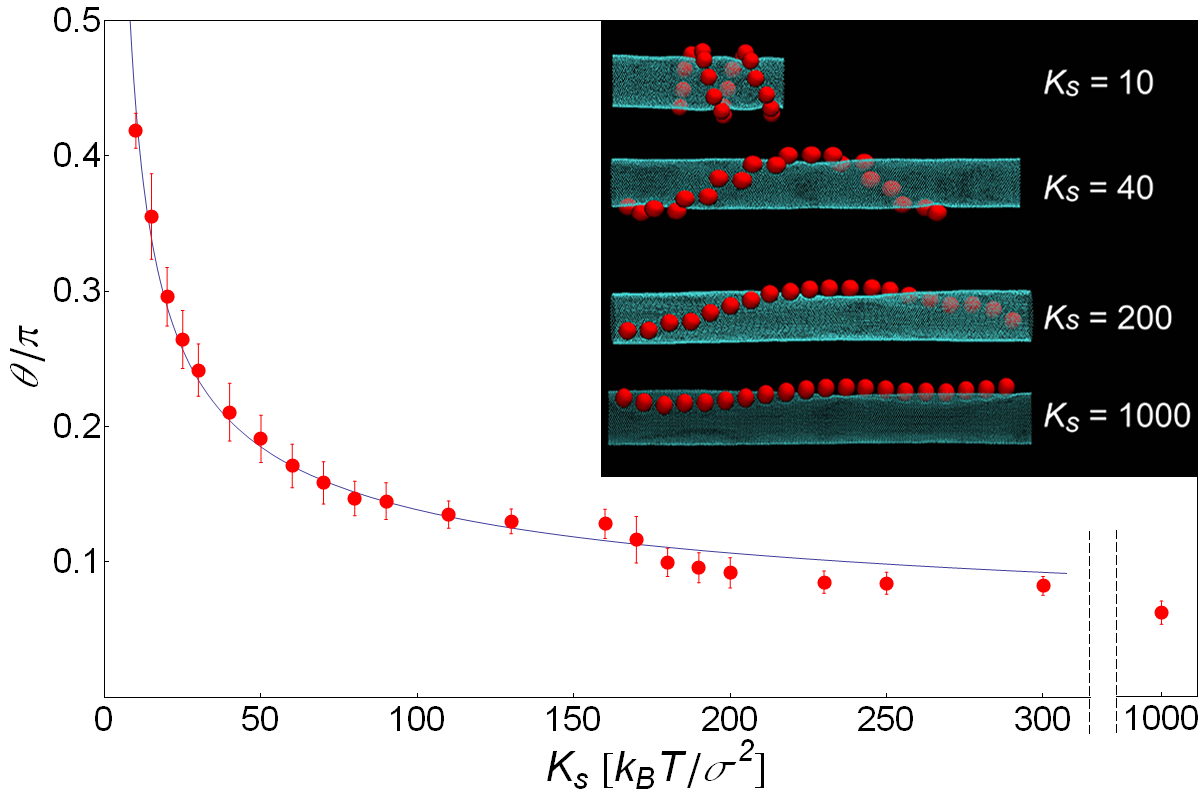}}
\subfigure[]{\label{fig:edge-b}\includegraphics[width=60mm]{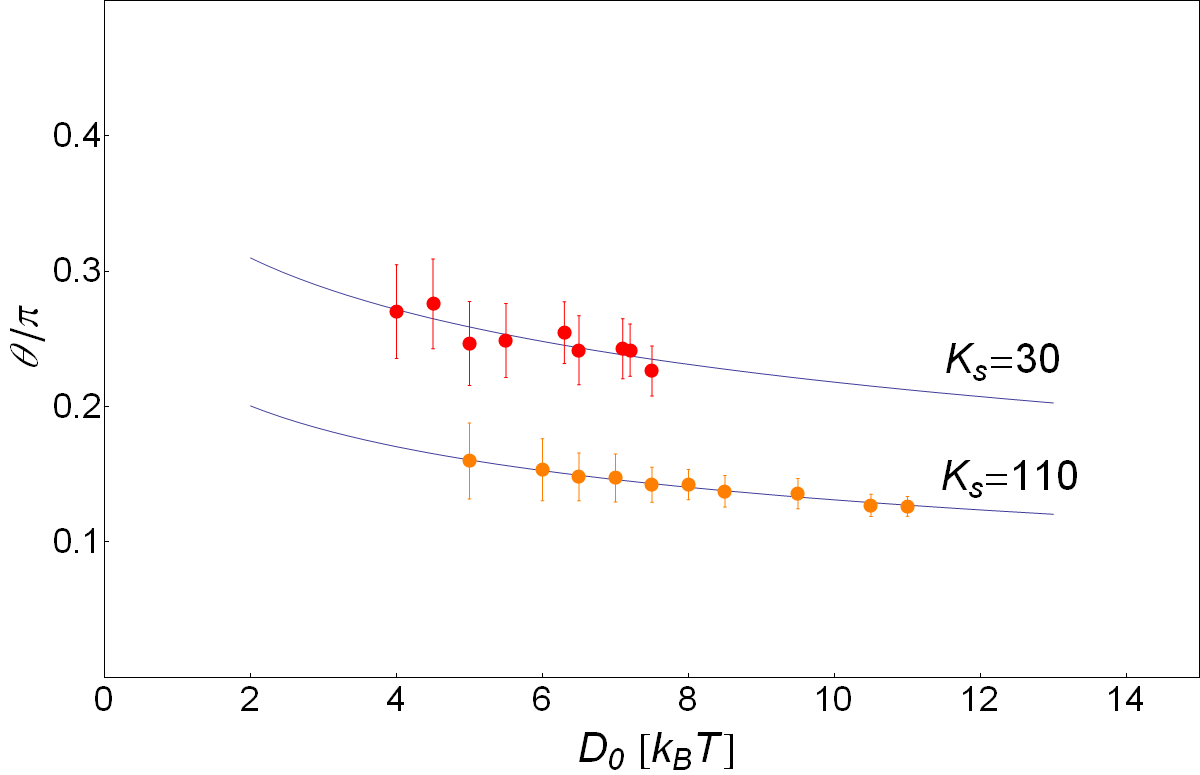}}
\caption{(a) Variation of $\theta$ as a function of $K_s$ at fixed $h \approx 1.17\sigma$ and $K_b=150k_{\rm B}T$. The solid line indicates the fit 
to the data using the inverse of Eq.(5). The inset shows the representative helices form increasing values of $K_s$. 
(b) Variation of $\theta$ as a function of the binding energy $D_0$, at $K_b=150k_{\rm B}T$, for two different values of the stretching constant: 
$K_s=30k_{\rm B}T/\sigma^2$ and $K_s=110k_{\rm B}T/\sigma^2$. The solid line indicates the fit 
to the data using the inverse of Eq.(5). }
\end{figure}
Two important points need to be emphasized. (1) The physical origin of the disordered-to-helix transition of the chain can be understood
in  terms of the usual balance between the entropy of the filament and the energy penalty associated with a random, non-optimal distribution of
indentations on the surface. (2) By going through the transition the filament acquires a large effective bending rigidity which results in a
persistence length several times larger than the chain length.

The jump in persistence length of the polymer can be best observed by measuring a function that accounts 
for the periodic correlation between the monomers, as described in~\cite{srebnik2}:
\begin{equation}
G(m)=\frac{1}{N_m-3}\displaystyle\sum_{i=1}^{N_m-2} g(m,i).
\label{G(m)}
\end{equation}
Here $m$ is the number of monomers between particle $i$ and $j$ along the chain, and $g(m,i)$ is given by
\begin{equation}
g(m,i)=\frac{(N_m-1)\sum_{j=1}^{N_m-m-1}(s_{i,j}-\overline{s_{i,j}})(s_{i,j+m}-\overline{s_{i,j}})}{(N_m-m-1)\sum_{j=1}^{N_m-1}(s_{i,j}-\overline{s_{i,j}})^2}
\label{g(m)}
\end{equation}
where $s_{i,j}=\cos\theta_{i,j}$ is the cosine of the angle between bond vectors $i$ and $j$, and $\overline{s_{i,j}}$ is the average over all such angles in the chain.
\begin{figure}[h!]
\center
\includegraphics[width=60mm]{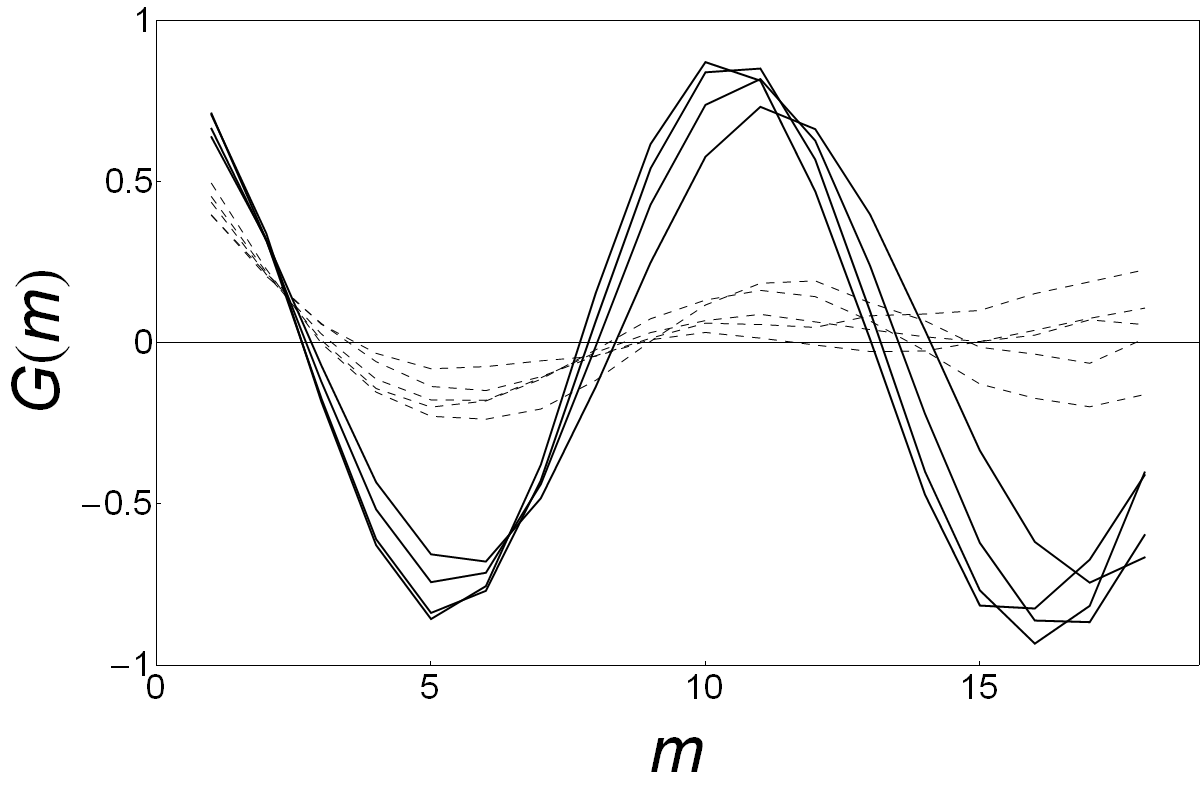}
\caption{ $G(m)$ as calculated from Eq.~(6) at $K_s=10k_{\rm B}T/\sigma^2$, $K_b=150k_{\rm B}T$, for different values of $D_0$. 
The jump in persistance length is observed around  $h \simeq 0.5 \sigma$. We indicate $G(m)$ with dashed lines for $h < 0.5 \sigma$ and we use solid lines
for $h>0.5\sigma$.}
\end{figure}
Figure 3 shows $G(m)$ for different values of $h$ at $K_s=10k_{\rm B}T/\sigma^2$ and $K_b=150k_{\rm B}T/\sigma^2$, and clearly indicates two distinct cases. For $h< 0.5\sigma$ the correlation between the relative location of the monomers on the surface is negligable, while for $h > 0.5\sigma$, $G(m)$ shows perfect helical correlation of monomers over a distance that is larger than $l$. Since $G(m)$ does not decay, it is obvious that our polymer is too short for a reliable estimate of the persistance length in the helical phase. 
However, the persistence length clearly exceeds the chain length over several times. 

What limits the length of the polymer in our simulations is the large number of triangles required to describe the cylindrical surface. In fact, to avoid the artifacts due to the specific tesselation of the surface, monomers need to be significantly larger than the surface beads. We find that the $\sigma_m=10\sigma$ is enough for the monomers not to feel the underlying structure of the membrane. Interestingly, when the monomer size becomes comparable to the size of the surface beads,
we find that the direction of the chain is biased along  the main axes of the mesh. This is a reminder
that below a certain length scale, the structural details  of the underlying surface
cannot be neglected.

In conclusion, it is important to emphasize two things. The first is that 
the onset indentation amplitude $h$ for helical conformation is
typically just a small fraction of the monomer size (barely 5\% in the case described above)
which is not an unreasonable perturbation even for simple membrane-bound proteins. The second is 
that although in this paper we have focused specifically on the problem of flexible chains on cylindrical surfaces,
the nature of this phenomenon is quite general and is intrinsically connected to the nonlinear response to
deformations of elastic sheets. This behavior can be generalized to arbitrary geometries $-$ we find that filaments also acquire
very peculiar conformations when placed on spherical or toroidal deformable shells~\cite{sphere} $-$ and more importantly 
can be extended to any component adhering to the surface. We can anticipate~\cite{pep}
that elastic surfaces can be used to drive self-assembly of hard colloidal particles resulting in
a variety of geometric patterns not unlike the ones observed with the filaments. Clearly the specific 
details of the long-range correlations induced by the surface will depend on the surface topology, and on the physical constraints
of the macromolecules adhering to it. Nevertheless, it is the interplay between the stretching and bending modes
of the surface that will determine the effective interactions between the components bound to it and the overall geometry of the aggregates.

Our hope is that the results presented in our paper will stimulate experimentalists to further study the elastic and mechanical properties 
of elastic sheets and, in particular the long range correlations arising when particles bind to it.

\section*{ACKNOWLEDGMENTS}
This work was supported by the National Science Foundation under Career Grant No. DMR-0846426.

\end{document}